\documentclass[epj]{webofc}
\usepackage[varg]{txfonts}
\usepackage{bm}
\usepackage{slashed}
\usepackage{overpic}
\usepackage{braket}
\woctitle{QCD@Work 2018}
\begin{document}
\title{Dual chiral density waves in nuclear matter}
\author{\firstname{Hiroaki} \lastname{Abuki}\inst{1}\fnsep%
\thanks{\email{abuki@auecc.aichi-edu.ac.jp}} \and
\firstname{Yusuke} \lastname{Takeda}\inst{2}\fnsep%
\thanks{\email{yusuketakeda4370@gmail.com}} \and
\firstname{Masayasu} \lastname{Harada}\inst{2}\fnsep%
\thanks{\email{harada.masayasu@nagoya-u.jp}}
}

\institute{Department of Education, Aichi University of Education,
1 Hirosawa, Kariya 448-8542, Japan \and
Department of Physics, Nagoya University, Nagoya, 464-8602, Japan}

\abstract{%
We study inhomogeneous chiral phases in nuclear matter using a hadronic
model with the parity doublet structure. 
With an extended ansatz for the dual chiral density wave off the chiral limit,
we numerically determine the phase structure. 
A new type of dual chiral density wave where the condensate has
nonvanishing space average is confirmed and it comes to occupy a wide
range of low density region as the chiral invariant mass parameter is lowered.
}
\maketitle

\section{Introduction}
The chiral symmetry breaking serves as a key ingredient to understand
the phase structure of QCD, since it is
responsible for the mass generation of hadrons as well as the mass
splittings of chiral partners. 
The chiral symmetry is expected to restore at sufficiently high baryon
density.
An interesting possibility opens up when one allows the chiral
condensate to vary in space; %
Nakano and Tatsumi demonstrated that the symmetry
restoration may take place via several steps \cite{Nakano:2004cd};
going up in density from the vacuum, the system first goes into an
intriguing state named as dual chiral density wave (DCDW), that is, 
a particular type of inhomogeneous chiral phases, making a spiral in the
$(\sigma_0,\pi_0)$ chiral plane along $z$ direction.

There are number of approaches to chiral inhomogeneous phases
\cite{Buballa:2014tba}.
These include; the mean-field approximation
\cite{Nickel:2009wj,Carignano:2010ac,Karasawa:2013zsa,Adhikari:2017ydi},
the Ginzburg-Landau expansion
\cite{Nickel:2009ke,Abuki:2011pf,Abuki:2013pla,Carignano:2017meb},
a self-consistent mean-field approach \cite{Lee:2017yea}.
These are based on quark-based models such as an NJL-type model.
One of the advantages is that these models are capable of realizing the
QCD vacuum properties as well as the color-flavor locked phase of quark
matter which is known to be the densest phase of QCD.
But there is also a disadvantage, that is, the lack of the ability to
reproduce properties of nuclear matter at saturation point, namely, the
QCD phase next to the vacuum phase being followed right after the
liquid-gas phase transition.

We here report on our recent work on this topic \cite{Takeda:2018ldi}
where we adopted a hadronic model with parity doublet picture (mirror
assignment) \cite{Detar:1988kn,Jido:2001nt}, with vector mesons included
in a manner guided by the hidden local symmetry
\cite{Bando:1987br,Harada:2003jx}.
We first extend the ansatz for the DCDW phase so as to take into account
the effect of the explicit symmetry breaking. 
The extended ansatz smoothly interpolates between the DCDW phase and a
nearly symmetry-restored phase.
With this setup, we construct the effective potential by diagonalizing
the  Bogoliubov-de Gennes (BdG) Hamiltonian for nucleons, and determine
phases via numerically minimizing the potential.
Our main finding is the emergence of another type of DCDW phase which
occupies the lower density region according to the value of chiral
invariant mass.

\section{Model}\label{sec:model}
In our analysis, we introduce $N^\ast(1535)$ as the chiral partner to
the ordinary nucleon based on the parity doublet structure.  
Along the line described in \cite{Motohiro:2015taa}, 
we construct a relativistic mean field model to describe nuclear
matter, which includes the scalar ($\sigma$), pseudoscalar ($\pi$)
mesons, and also the vector ($\omega$) meson within the unitary gauge of the hidden local symmetry.

The pure mesonic part for the Lagrangian is then given by
\begin{equation}
\begin{array}{rcl}
 {\cal
  L}_{\mathrm{mes.}}&=&\frac{m_\omega^2}{2}\omega_\mu\omega^\mu-\frac{1}{4}\omega_{\mu\nu}\omega^{\mu\nu}%
 +\frac{1}{2}\partial_\mu\sigma\partial^\mu\sigma+\frac{1}{2}\partial_\mu\bm{\pi}\partial^\mu\bm{\pi}\\[2ex]
&&+\frac{\bar{\mu}^2}{2}(\sigma^2+\bm{\pi}^2)-\frac{\lambda_4}{4}%
(\sigma^2+\bm{\pi}^2)^2%
  +\frac{\lambda_6}{6}(\sigma^2+\bm{\pi}^2)^3-f_\pi m_\pi^2\sigma,
\end{array}
\end{equation}  
where $\bar{\mu}^2$, $\lambda_4$ and $\lambda_6$ are model parameters.
$\omega_\mu$ and $\omega_{\mu\nu} = \partial_\mu \omega_\nu -
\partial_\nu \omega_\mu$ are the field and field strength for $\omega$
meson, which are singlet under the chiral transformation. 
The mass and the decay constant of the pion, as well as the mass of
omega meson, are to be set to physical values $m_\pi=139$~MeV,
$f_\pi=92.2$~MeV, and $m_\omega=783$~MeV, respectively.

In the present model based on the parity doublet structure,  
the transformation properties of the positive and negative parity nucleon fields are given by
\begin{equation}
\begin{array}{rclrcl}
\psi_{1r}&\to& g_{R}\psi_{1r},\quad&\psi_{2r}&\to& g_{L}\psi_{2r},\\
\psi_{1l}&\to& g_{L}\psi_{1l},\quad&\psi_{2l}&\to& g_{R}\psi_{2l},
\end{array}
\end{equation}
where $g_{R}$ ($g_{L}$) is an element of SU(2)$_{R}$ (SU(2)$_{L}$) chiral symmetry group, and $\psi_{1r}$ and $\psi_{2r}$ ($\psi_{1l}$ and $\psi_{2l}$) are
the right-handed (left-handed) fields.
Based on the transformation properties, the Lagrangian relevant for
nucleons is expressed as~\cite{Motohiro:2015taa}
\begin{equation}
\begin{array}{rcl}
 {\cal
  L}_{\mathrm{N}-\mathrm{mes}}&=&\,\bar\psi_{1}(i\slashed{\partial}+\slashed{\mu}_B%
    -g_\omega\slashed{\omega})\psi_{1}%
   +\bar\psi_{2}(i\slashed{\partial}+\slashed{\mu}_B-g_\omega\slashed{\omega})\psi_{2}%
   -m_0\left(\bar{\psi}_2\gamma_5\psi_1-\bar{\psi}_1\gamma_5\psi_2\right)\\[2ex]
           && -g_1\bar\psi_{1}(\sigma+i\gamma_5\bm{\pi}\cdot\bm{\tau})\psi_{1}%
            -g_2\bar\psi_{2}(\sigma-i\gamma_5\bm{\pi}\cdot\bm{\tau})\psi_{2},
\end{array}
\label{eq:LN}
\end{equation}
where the baryon chemical potential is included via $\slashed{\mu}_B=\mu_B\gamma_0$.
$\bm{\tau}=(\tau^1,\tau^2,\tau^3)$ is a set of Pauli matrices, $g_1$,
$g_2$ and $g_{\omega}$ are the coupling constants. 
$m_0$ is the chiral invariant mass parameter which survives as a
nonvanishing nucleon mass even when the chiral symmetry is restored.

In the present analysis, we adopt the following extended DCDW ansatz,
$\sigma=\delta\sigma+\sigma_0\cos(2fz)$, $\pi_3=\sigma_0\sin(2fz)$, that
reads in the complex representation:
\begin{equation}
 \langle\sigma+i\bm{\pi}\cdot\bm{\tau}\rangle=\delta\sigma+\sigma_0e^{2ifz\tau^3}
\equiv M(z),
\label{eq:ansatz}
\end{equation}
where $\delta \sigma$, $\sigma_0$ and $f$ are variational parameters
with dimension one.
Space independent part $\delta\sigma$ accommodates the possibility that
the space average of DCDW condensate would get nonvanishing shift into
$\sigma$-direction due to the explicit chiral symmetry breaking.
In the mean-field approximation, the nucleon contribution to the
effective potential can be written as
\begin{align}
 \Omega_{N}=\frac{i}{V_4}\mathrm{Tr}\,%
 \mathrm{Log}(i\partial_0-({\mathcal H}(z)-\mu_B^*)),
\label{eq:omegaB}
\end{align}
where $V_4$ is the space-time volume and ${\mathcal H}(z)$ is the single
particle Bogoliubov-de Gennes (BdG) Hamiltonian defined in the space of
fermion bispinor $\psi=(\psi_1,\psi_2)$ as
\begin{equation*}
{\mathcal H}=\left(%
\begin{array}{cc}
i\gamma^0\bm{\gamma}\cdot\nabla+g_1\gamma^0(M(z)^\dagger P_L+M(z)P_R) & m_{0}\gamma^0(P_R-P_L) \\
m_{0}\gamma^0(P_L-P_R) &
 i\gamma^0\bm{\gamma}\cdot\nabla+g_2\gamma^0(M(z)P_L+M(z)^\dagger P_R) \\
\end{array}
\right),
\end{equation*} 
where $P_{R(L)}=\frac{1\pm\gamma_5}{2}$ is the chirality projector.
This is nothing but the Dirac Hamiltonian in the presence of a periodic
potential field $M(z)$ ($=M(z+\lambda)$ with a wavelength $\lambda=\pi/f$). 
Then the functional trace in Eq.~(\ref{eq:omegaB}) can be evaluated by
finding eigenvalues of the operator ${\mathcal H}(z)$
\cite{Nickel:2008ng}.
The eigenvalue has a discrete label as well as continuous three-momentum
${\bf p}$ in addition to internal quantum numbers; 
This is because of the Bloch theorem which states that the
eigenfunctions in the presence of a periodic potential are the plane
waves distorted by periodic functions.
To be specific, we decompose the bispinor as
\begin{equation*}
\psi({\bf x})=\sum_{\ell=-\infty}^{\infty}%
\sum_{{\bf p}_\perp}\sum_{|p_z|\le f}%
e^{i{\bf p}\cdot{\bf x}+i{\bm{K}}_\ell\cdot{\bf x}}\psi_{{\bf p},\ell}\,,
\end{equation*}
where ${\bm{K}}_\ell=(0,0,2f\ell)$ is the reciprocal lattice vector.
Moving on to the quasimomentum base $\{\psi_{{\bf p},\ell}\}$, the BdG
Hamiltonian becomes block-diagonalized, 
$H_{\ell\ell'}({\bf p})$.
Since the isospin remains a good quantum number, we can simply double
the proton contribution in the full effective potential.
Then omitting the antiprotons which would not contribute at zero
temperature, and diagonalizing $H_{\ell\ell^\prime}({\bf p})$ results in
an infinite tower of eigenvalues at each ${\bf p}$, which 
repeatedly appears for every Brillouin %
Zone (BZ), ${\bf p}\to {\bf p}+\bm{K}_\ell$ ($\ell=\cdots,-1,0,1,\cdots$):
\begin{equation*}
\sum_{\ell^\prime}%
 H_{\ell\ell'}({\bf p})\psi^{(i)}_{n,{\bf p},\ell^\prime}%
 =E_{n,\bf{p}}^{(i)}\psi^{(i)}_{n,\bf{p},\ell}\,%
 \quad(n=0,1,\cdots,\infty),
\end{equation*}
with $i(=1, 2, 3, 4)$ labeling the internal quantum number
$(p,p^*)\otimes(\uparrow,\downarrow)$, where $p^\ast$ implies the
$I_3=+1/2$ part of $N^\ast(1535)$.
The total thermodynamic potential is now evaluated as
\begin{equation}
\begin{array}{rcl}
\Omega&=&\displaystyle\sum_{n=0}^{\infty}%
 \sum_{i=1}^4\int_{-f}^f\frac{dp_z}{\pi}%
 \!\!\int\frac{d{\bf p}_\perp}{(2\pi)^2}%
 (E^{(i)}_{n,{\bf p}}-\mu_B^*)\theta(\mu_B^*-E^{(i)}_{n,{\bf p}})\\
  &&-\frac{1}{2}m_\omega^2\omega_0^2%
           -\frac{1}{2}\bar\mu^2\left(%
           \delta\sigma^2+\sigma_0^2%
           \right)+2\sigma_0^2f^2%
   +\frac{1}{4}\lambda_4\left[%
           \left(\delta\sigma^2+\sigma_0^2\right)^2%
           +2\delta\sigma^2\sigma_0^2%
           \right]\\[2ex]
  &&-\frac{1}{6}\lambda_6\left[%
           \left(\delta\sigma^2+\sigma_0^2\right)^3%
           +6\left(\delta\sigma^2+\sigma_0^2\right)%
           \delta\sigma^2\sigma_0^2\right]-m_\pi^2f_\pi\delta\sigma.
\end{array}
\label{eq:Omega}
\end{equation}

\begin{table}[b]
\caption{Determined parameters for given chiral invariant mass}
\label{table:model-para}
\centering
\begin{tabular}{c|cccccc}
\hline\hline
\ $m_{0}$ \ & \ $500$ \ & \ $600$ \ & \ $700$ \ & \ $800$ \ & 900 \ \\
\hline
$g_1$&$9.03$&8.49&7.82&7.00&5.97\\
$g_2$&$15.5$&15.0&14.3&13.5&12.4\\
$g_{\omega}$&11.3&9.13&7.30&5.66&3.52\\
$\bar\mu\,[\rm{MeV}]$&441&437&406&320&114\\
$\lambda_4$&42.2&40.6&35.7&23.2&4.47\\
$\lambda_6\cdot f_\pi^2$&17.0&15.8&14.0&8.94&0.644\\
\hline\hline
\end{tabular}
\end{table}

\noindent
\paragraph{\bf Parameter setting}
We treat $m_0$ as a free parameter.
Parameters $g_1$ and $g_2$ can be determined by QCD vacuum property.
Once $m_0$ is given, $m_+^{(0)}=939$~MeV and
$m_-^{(0)}=1535$~MeV together with $\sigma_0=f_\pi=92.2$~MeV fix the values of
$g_1$ and $g_2$ via
\begin{align}
 m_{\pm}^{(0)}=\frac{1}{2}\left[%
   \sqrt{\left(g_1+g_2\right)^2\sigma_0^2+4m_0^2}%
   \mp\left(g_1-g_2\right)\sigma_0%
   \right].
\end{align}
From the stationary condition for $\sigma$ at vacuum, we have 
$\bar\mu^2=\lambda_4f_\pi^2-\lambda_6f_\pi^4-m_\pi^2$.
Then we are left with three unknown parameters to be fixed,
$\{\lambda_4,\lambda_6,g_{\omega}\}$.
In order to fix these parameters, we use nuclear matter property
at saturation density as done in \cite{Motohiro:2015taa,Takeda:2017mrm}:
\begin{equation}
\begin{array}{rcl}
\rho_0&=&0.16~\mbox{fm}^{-3},\\[2ex]
\displaystyle
\left(%
\frac{\epsilon}{\rho}\right)\Bigg|_{\rho=\rho_0}-m_+^{(0)}%
=\mu_B^*|_{\rho=\rho_0}-m_+^{(0)}&=&-16~\mbox{MeV},\\[3ex]
K=\displaystyle9\rho_0^2\frac{\partial^2(\epsilon/\rho)}{\partial^2\rho}%
\bigg|_{\rho=\rho_0}%
=9\rho_0\frac{\partial P}{\partial\rho}\bigg|_{\rho=\rho_0}%
&=&240~\mbox{MeV}.
\end{array}
\label{eq:conditions}
\end{equation}
The first and second equations determine the value of $g_{\omega}$, and
$\omega_0$ as a function of density, $\omega_0=g_{\omega}\rho/m_\omega^2$.
The saturation condition $\frac{\partial}{\partial\rho}%
(\epsilon/\rho)\big|_{\rho=\rho_0}=P|_{\rho=\rho_0}/\rho_0^2=0$, 
and the condition for the incompressibility, the last equation of
Eq.~(\ref{eq:conditions}), 
together with the stationary condition
 $\frac{\partial\Omega}{\partial\sigma}|_{\rho=\rho_0}=0$
determines parameters
$\{\lambda_4,\lambda_6\}$ and the scalar condensate at saturation
density, $\sigma_0|_{\rho=\rho_0}$.
The parameters are summarized in Table \ref{table:model-para}.
As will be shown later, however, the saturated nuclear matter exists only
as a metastable state once chiral invariant mass becomes smaller than
some critical value, $m_{0}\lesssim 800$~MeV.


\begin{figure}[tbp]
\begin{center}
\includegraphics[scale=0.8,clip]{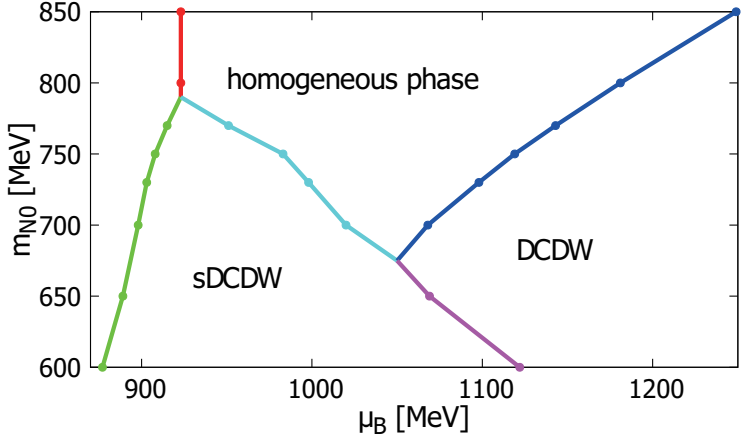}
\caption{Phase structure in $\mu_B-m_{0}$ plane. 
Red line is $\mu_B=923$~MeV corresponding to the nuclear liquid-gas
 phase transition point.}
\label{fig:1}
\end{center}
\end{figure}
\section{phase structure}\label{sec:PhaseStructure}
Figure~\ref{fig:1} shows the phase structure in $\mu_B-m_{0}$ plane.
We notice that there are two kinds of DCDW phase, ``DCDW'' and ``sDCDW''.
Depending on the range where $m_0$ resides, qualitative picture of phase
transitions changes.

\begin{figure}[tb]
\centering
\begin{overpic}[scale=0.5,clip]{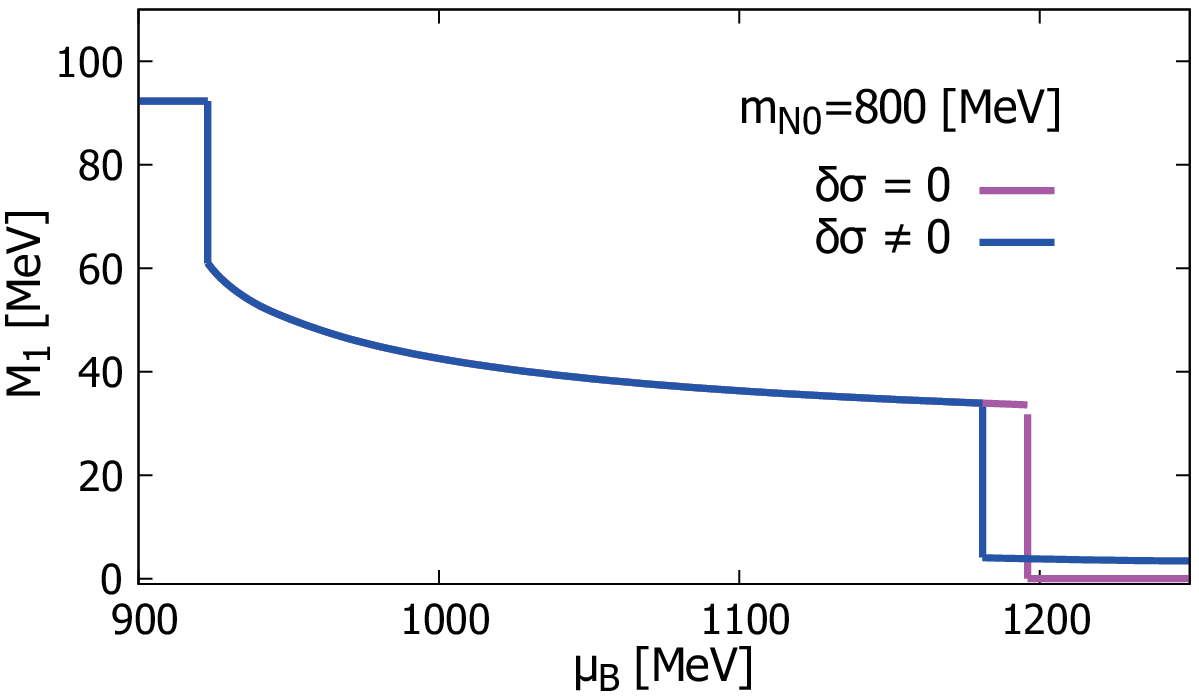}
\put(50,60){(a)}
\end{overpic}
\hspace*{0.05ex}
\begin{overpic}[scale=0.5,clip]{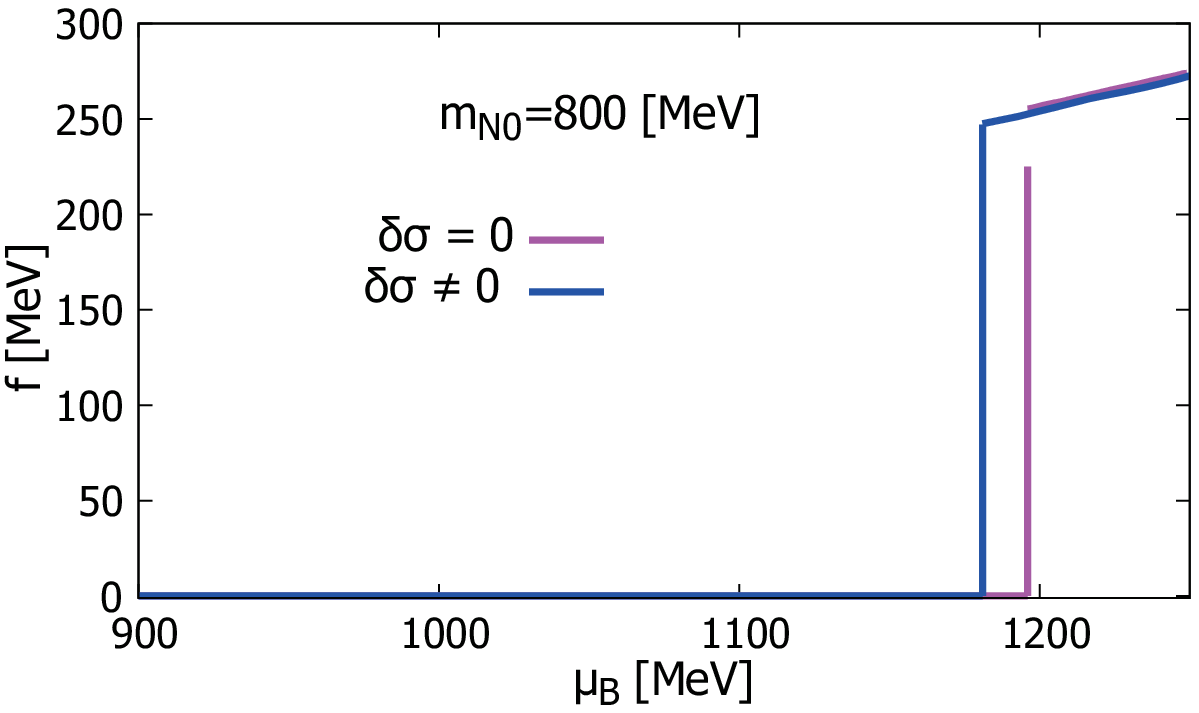}
\put(50,60){(b)}
\end{overpic}\\
\vspace*{3.5ex}
\hspace*{1.15ex}
\begin{overpic}[scale=0.5,clip]{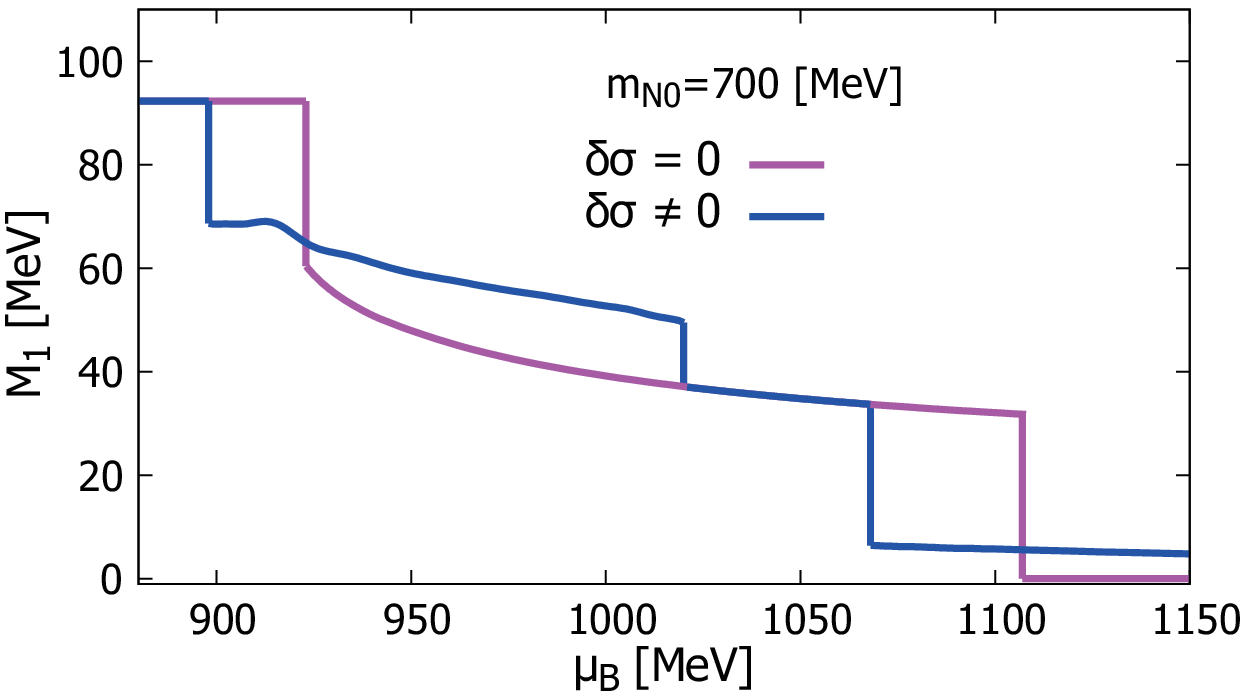}
\put(48,58){(c)}
\end{overpic}
\hspace*{-1.6ex}
\begin{overpic}[scale=0.5,clip]{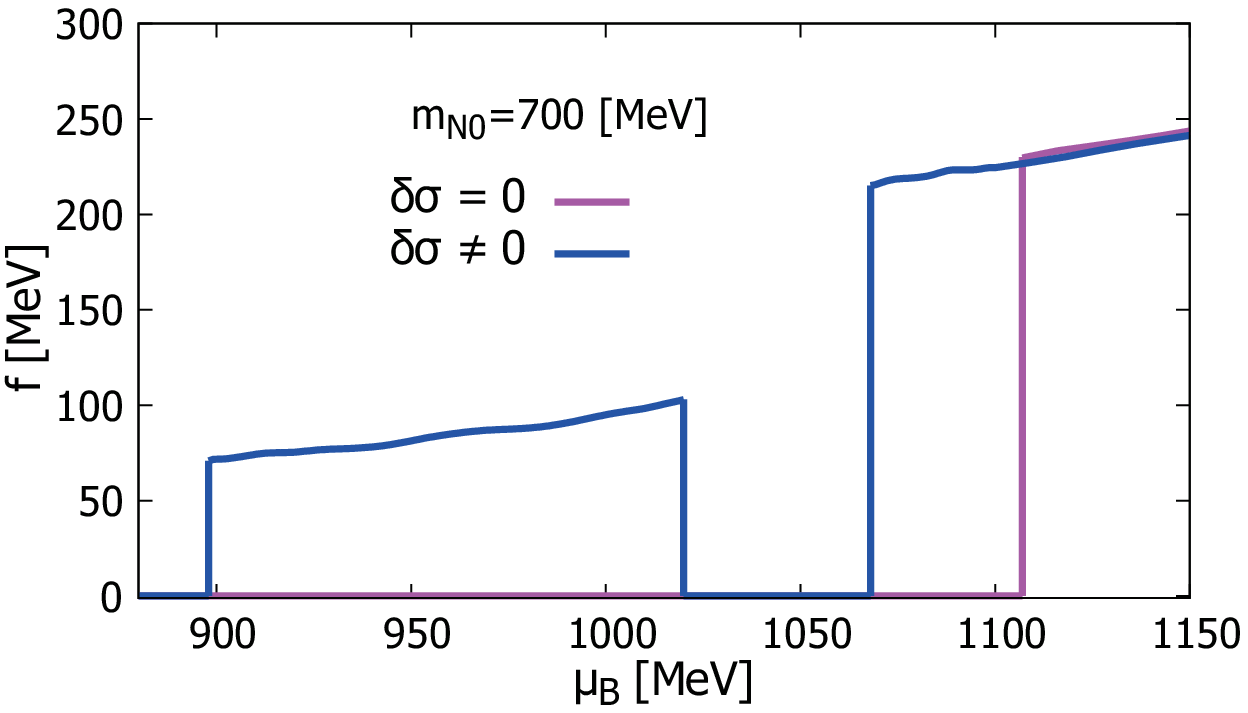}
\put(48,58){(d)}
\end{overpic}
\caption{
Left:~the order parameter $M_1$ (see text) as a function of $\mu_B$ for
 $m_0=800$~MeV (upper panel) and that for $m_0=700$~MeV (lower panel).
Right:~the wavenumber $f$ as a function of $\mu_B$ for $m_0=800$~MeV
(upper panel) and that for $m_0=700$~MeV (lower panel).
Magenta curves correspond to the solutions obtained under the forced
 condition $\delta\sigma=0$.
}
\label{fig:2}
\end{figure}

For $m_0\gtrsim780$~MeV, there are three phases; (1)~the vacuum
phase for $\mu_B\le 923$~MeV, (2)~the homogeneously chiral symmetry
broken phase, and (3)~the DCDW phase at high density.
In the upper panel of Fig.~\ref{fig:2}, we show order parameters as a
function of $\mu_B$ for $m_0=800$~MeV.
Just for comparison we also depict by magenta curve, the solution for
the case where the condition $\delta\sigma=0$ is forced (that is, the
case of the standard DCDW ansatz).
Plotted in Fig.~\ref{fig:2}(a) is the space-averaged order parameter
$M_1$, defined by
\begin{equation}
 M_1\equiv\frac{1}{2V}\int_{-\infty}^{\infty}d^3x%
           {\rm tr}\left[\braket{M}\right]=\begin{cases}
	    \delta\sigma+\sigma_0 & (f=0), \\
	    \delta\sigma & (f\neq0),\\
	   \end{cases}
 \label{eq:M1}
\end{equation}
which provides a guide for the strength of chiral symmetry
breaking.
Figure \ref{fig:2}(b) shows wavenumber $f$ as a function of $\mu_B$.
Nonvanishing wavenumber ($f\ne0$) means the DCDW phase where the
translational symmetry is also broken.
We note that, when a shift $\delta\sigma$ is properly taken into
account, the onset density the DCDW phase is brought to a lower density
since it stabilizes the DCDW solution through the explicit symmetry
breaking source, $-f_\pi m_\pi^2\delta\sigma$.

For $670~\mbox{MeV}\lesssim m_0\lesssim 780~\mbox{MeV}$, we find four phases. 
These are, going up in density, (1) the vacuum phase, (2)
the ``sDCDW'' phase, (3) the homogeneously chiral-symmetry broken phase, and
(4)~the DCDW phase.
What we call by ``sDCDW'' phase is the ``shifted'' DCDW phase, for a
reason we will describe shortly.
In order to see how phase transitions take place, we display in the
lower panel of Fig.~\ref{fig:2}, the order parameters as a
function of $\mu_B$ for $m_0=700$~MeV.
From these we see that, as the baryon density increases, the chiral
symmetry restores through several steps. 
Moreover, we clearly see from the Fig.~\ref{fig:2}(d), that there are
two DCDW phases; `one at low density, and the other at
high density, being separated by the homogeneously chiral-symmetry
broken phase. 
We call the former ``sDCDW'', and the latter ``ordinary'' DCDW or simply
DCDW.

Let us now have a closer look at the sDCDW which shows up next 
to the vacuum phase, in the region
$900~\mathrm{MeV}\lesssim\mu_B\lesssim1020~\mathrm{MeV}$ for $m_0=700$~MeV.
In contrast to the ordinary DCDW for $\mu_B\gtrsim1070$~MeV
the phase is not smoothly connected to the DCDW phase found at high
density side for $m_0=800$~MeV as clearly seen in Fig.~\ref{fig:1}.
In the ordinary DCDW, the magnitude of the wavenumber $f$ is
much larger than $\delta\sigma$ (and also than $\sigma_0$).
The chiral density wavelength is roughly $\lambda=\pi/f\sim 2$ --
$3$~fm
which is still larger than averaged inter-nucleon spacing
$\rho^{-1/3}\sim 1.2$ -- $1.3$~fm.
In the sDCDW phase, on the other hand, the wavenumber $f\sim 70$
--$100$~MeV is comparable with $\delta\sigma\sim 50$ -- $70$~MeV.
The resulting chiral density wavelength $\lambda=\pi/f\sim6$ -- $9$~fm,
which is much larger than the averaged inter-nucleon spacing
$\rho^{-1/3}\sim 1$ -- $2$~fm.
Most interesting fact is that in the sDCDW phase the amplitude of
condensate is smaller than the magnitude of the shift, namely,
$\sigma_0<\delta\sigma$.
Accordingly, the center of the chiral spiral in the $(\sigma,\pi^0)$
chiral plane, which is located near origin in the ordinary DCDW phase, 
is significantly shifted to the $\sigma$ direction.
This is why we name the phase the ``shifted'' DCDW (abbreviated to
``sDCDW'') phase.
We would like to stress that, on the contrary to the ordinary DCDW
phase, $\delta\sigma$ in sDCDW phase is not due to the symmetry breaking
source term, but rather spontaneously generated.
This means, the solution to the stationary condition for $\delta\sigma$
would not vanish even in chiral limit.

As the chiral invariant mass $m_0$ is decreased, the density window for
the  homogeneously chiral-symmetry broken phase shrinks as the pressure
of two kinds of DCDW phase gets stronger.
And once the condition $m_0\lesssim670$~MeV is met, it does no longer exist.

\section{Conclusions and outlook}\label{sec:summary}
We studied the inhomogeneous phase structure in nuclear matter using a
nucleon-based model with parity doublet structure where $N^\ast(1535)$
is introduced as the chiral partner of $N(939)$.
Adopting an extended ansatz for DCDW, Eq.~(\ref{eq:ansatz}), we found
that, the sDCDW phase shows up in addition to the ordinary DCDW phase
when the value of chiral invariant mass $m_0$ is below some threshold,
$\sim780$~MeV.

The ordinary DCDW phase appears at high density, where the space average
of chiral condensate $M_1$, Eq.~(\ref{eq:M1}), is typically less than 10~MeV.
This implies that this phase smoothly transforms into the standard DCDW
with $\delta\sigma=0$ as the chiral limit is approached.
The wavenumber $f$ has value of $200\sim300$~MeV, corresponding to the
density wavelength $\lambda\sim2$ -- $3$~fm, which is in fair agreement
with the result obtained in \cite{Heinz:2013hza}.

On the other hand, when $m_{0}\lesssim780$~MeV, the sDCDW phase appears
at low density. 
This phase is characterized by a smaller wavenumber $f$ and a large
shift of chiral condensate, $\delta\sigma$.
It is noteworthy that it is not the effect of explicit chiral
symmetry breaking but the dynamical symmetry breaking that produces this
large shift of chiral condensate. 
So we expect that this sDCDW phase survives in the chiral limit.

The parameter range of chiral invariant mass where the sDCDW is
stabilized, fails to realize nuclear matter as the pressure of sDCDW is
so strong that it washes out the liquid-gas phase transition structure.
Then, one might think that the present model for $m_{0}$ less than
780~MeV is ruled out.
However, the chiral invariant mass $m_{0}$ can have density dependence
as in Ref.~\cite{Heinz:2013hza} which shows that $m_{0}$ decreases
against increasing density.  In such a case, the sDCDW phase may be
realized in high density nuclear matter in the real world.

Exploring the elementary excitations in the sDCDW phase deserves further
investigations in future.
In the ordinary DCDW phase, a particular combination of the space
translation and chiral rotation remains unbroken
\cite{Lee:2015bva,Hidaka:2015xza}.
As a result, there is no extra Nambu-Goldstone boson other than
three pions.
In contrast, there is no such unbroken combination of symmetry
in the sDCDW phase.
We then expect that a phonon mode appears in the sDCDW phase which may
signal the phase.
An inclusion of external magnetic field may provide another interesting
direction of extending current work.
For quark matter, several studies were already devoted to the topic of
inhomogeneous phases under the magnetic field
\cite{Frolov:2010wn,Nishiyama:2015fba,Yoshiike:2015tha,%
Cao:2016fby,Abuki:2018iqp}.
The analysis within the nucleon-based model with
including both $\delta\sigma$ and magnetic field would be an interesting
subject worth exploring.

\smallskip
\noindent
{\it Acknowledgement.}~This work was partially supported by JPSP KAKENHI
Grant Number JP16K05346 (H.A.) and 16K05345 (M.H.).

\end{document}